\documentclass[journal]{IEEEtran}
\ifCLASSINFOpdf
\else
   \usepackage[dvips]{graphicx}
\fi
\usepackage{comment}
\usepackage{cite}
\usepackage{times,epsfig,cite,latexsym,fancybox,amsmath,amssymb}
\usepackage{graphicx,cite,epsfig,amssymb,amsmath,endnotes,algorithm,algorithmic}
\usepackage{algorithmic}
\usepackage{graphicx}
\usepackage{textcomp}
\usepackage{xcolor}
\usepackage[utf8]{inputenc}
\usepackage{multirow}
\usepackage{float}
\usepackage{physics}
\usepackage{amsmath,amssymb}
\DeclareMathOperator{\E}{\mathbb{E}}
\usepackage{enumitem}
\usepackage{amsmath}
\usepackage{textcomp}
\usepackage{url}
\usepackage{comment}
\def\BibTeX{{\rm B\kern-.05em{\sc i\kern-.025em b}\kern-.08em
    T\kern-.1667em\lower.7ex\hbox{E}\kern-.125emX}}

\newtheorem{claim}{Claim}

\newcommand{\paren}[1]{\left({#1}\right)}

\usepackage{physics}

\usepackage{physics}
\renewenvironment{cases}{\left\{\begin{array}[c]{ll}}{\end{array}\right.}
\begin{document}
\title{ Low Complexity Single Source DOA Estimation Based on Reduced Dimension SVR}

\author{Md Imrul Hasan, \IEEEmembership{Student Member, IEEE}, and Mohammad Saquib, \IEEEmembership{Senior Member, IEEE}}
\markboth{Journal of \LaTeX\ Class Files, Vol. x, No. x, X 2021}
{Shell \MakeLowercase{\textit{et al.}}: Bare Demo of IEEEtran.cls for IEEE Journals}
\maketitle
\vspace{-20pt}
\begin{abstract}
Conventional direction of arrival (DOA) estimation algorithms suffer from performance degradation due to antenna pattern distortion and substantial computational complexity in real-time execution. The support vector regression (SVR) approach is a possible solution to overcome those limitations. In this work, we propose a sequential DOA estimation technique that combines the reduced dimension SVR (for the azimuthal plane) with a closed form approach (for the elevation plane). Thus, the training and testing are only required for the azimuthal angles which makes it very attractive from the implementation complexity point of view. Our analysis demonstrates that the proposed algorithm offers significant complexity gain over the popular MUSIC algorithm while exhibiting similar root-mean-square error performance.


\end{abstract}
\IEEEpeerreviewmaketitle
\vspace{-5pt}
\section{Introduction}
\IEEEPARstart{D}{irection} of arrival (DOA) plays a significant role in numerous fields such as sonar, radar,  navigation, geophysics, wireless communication, acoustic tracking, astronomy, defense operations, etc \cite{a01, a05, a07, a08,b1, b2, hasan2021rectangular }. Over several decades, many methods have been proposed to estimate the DOAs, i.e., multiple signal classification (MUSIC) \cite{c1}, maximum-likelihood (ML) \cite{c2}, capon \cite{c3}, estimation of signal parameters via rotational invariance techniques (ESPRIT) \cite{c4}, Min-Norm \cite{c5}, etc. However, these algorithms have two main disadvantages. Firstly, they are prone to a considerable computational burden in real-time applications, and secondly, malfunction of even one element causes severe degradation in the estimation performance as these algorithms assume all the antenna elements have the same response \cite{raj2007determination}.
 
 An alternative approach to these algorithms is the use of neural networks\cite{r2}. These networks offer offline training advantages and faster real-time computation of the DOAs than the conventional algorithms. However, those can produce multiple solutions while optimizing the cost function, or they may be subject to over-fitting due to the complexity of their structures. These issues have been resolved by the introduction of the support vector regression (SVR) approach \cite{a1, a3}. The SVR approximates  the input-output relationship through a non-linear mapping function where the parameters are computed during the training phase. The SVR based approaches provide several advantages;
\begin{enumerate}
    \item These methods are robust to errors due to antenna pattern distortion and do not require the prior information of the antenna pattern data.
    \item  These are easy to implement, very simple to operate, and provide a quick real-time estimation of the DOAs.
\end{enumerate}
 While implementing the SVR in DOA estimation, most of the works in the literature use the autocorrelation matrix elements of the received signal as their input data \cite{a1, a3}. In \cite{a005}, the ratio of the amplitudes, and phases from two received signals in two different antennas are considered as the input vector. Here, if the value in the denominator is close to zero, this approach could yield unreliable angle estimates. Importantly, the above SVR based works deal with only the 1D estimation of the DOAs (i.e., the estimates of the elevation angles).

 In this paper, we illustrate how to localize a single narrow-band source by combining the SVR with a closed-form method. We develop and analyze a reduced dimension SVR algorithm where both the training and testing are performed only for the azimuthal angles. The estimate of the azimuthal angle is then utilized to obtain the elevation angle using a closed-form approach, namely CFA. From now on, we refer the proposed algorithm to SVR-CFA. 
 
The contributions of this correspondence item can be outlined as follows:
\begin{enumerate}
\item  A 2D DOA training and testing are converted into a 1D problem, which greatly reduces the implementation complexities.

\item  In SVR-CFA, the phases of the autocorrelation of the received signals are proposed to use as the input data. 

\item When there is a chance of bust in the input data, that information is exploited immediately to decide the DOA.

\item The SVR-CFA is shown to notably outperform the popular MUSIC algorithm in terms of complexity while exhibiting almost identical root-mean-square error (RMSE) performance.
\end{enumerate}

\textit{Notations:} We use lowercase bold letters to denote vectors, uppercase bold letters to symbolize matrix, and lowercase letters in italics to represent scalars. The notation * refers to the complex conjugate, and $[\cdot]^\dag$ denotes the Hermitian of a matrix. The $\Vert\cdot\Vert$ notation is used to represent the norm of a vector, whereas $|\cdot|$ stands for the absolute value of a scalar. Lastly, $\angle \cdot$ signifies the angle of a complex number, and $\langle \cdot , \cdot \rangle$ is used to denote the scalar product between two vectors.

\section{Mathematical Formulation}
\label{s1}
The implantation of the proposed SVR-CFA does not hinge on the type of the planar arrays. However, for the sake of demonstration, we choose a uniform circular antenna (UCA) array. This array is extensively studied in the literature due to its advantage of providing azimuthal coverage as well as the elevation information of the DOAs \cite{lc1,lc2}. 
\begin{figure}[t!]
\centering
    \includegraphics[scale=0.65]{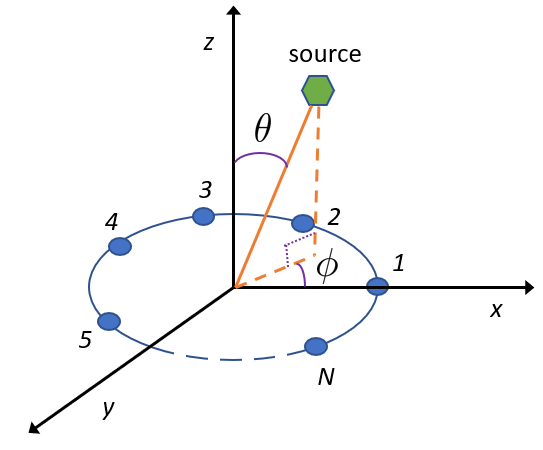}
    \caption{UCA array geometry.}
    \label{fig:Capture1}
\end{figure} 

Let's consider a UCA array of radius $r$ with $N$ antenna elements which are positioned on the circumference of the circle along the $xy-$plane. The first antenna is located on the $x-$axis, and the remaining antennas are placed in the anti-clockwise direction as shown in Fig.~\ref{fig:Capture1}. At the $m^{\underline{\mathrm{th}}}$ snapshot, the array is excited by a complex single narrow band wide sense stationary source signal, $s(m)$ with average power ${\sigma_\mathrm{s}}^2$. The azimuthal angle, $\phi$ of this signal is measured counterclockwise from the $x-$axis, and the elevation angle, $\theta$ is measured downward from the $z-$axis. 

For the UCA, the steering term corresponding to the $n^{\underline{\mathrm{th}}}$ antenna element can be expressed as
\begin{equation}
    {a_{{s}_n}}=e^{j\frac{2\pi}{\lambda}r\sin\theta \cos(\phi-\beta_n)},
    \label{a_s}
\end{equation}
where $\beta_n ={2\pi(n-1)}/{N}$; $n=1,2,...,N$, and $\lambda$ denotes the wavelength. Hence, the output of sensor $n$ is
  \begin{equation}
 {{ { {x}}}_n}(m)= {a_{s_n}}s(m)+w_n(m),
 \label{a_n0}
 \end{equation}
 where $w_n(m)$ is zero-mean white complex Gaussian noise with average power ${\sigma}^2$, and spatiotemporally independent of the source signal, $s(m)$. Now collecting outputs from all the elements, we form the received signal vector,
   \begin{equation}
    \mathbf{{x}}(m)=\mathbf{a}{s(m)}+\mathbf{w}(m).
    \label{RSE}
 \end{equation}
Here, 
 \begin{align*}
 \mathbf{{x}}(m)=\begin{bmatrix}
{x}_1(m)\\
{x}_2(m)\\
\vdots\\
{x}_N(m)
\end{bmatrix};
 \mathbf {a}=\begin{bmatrix}
a_{s_{1}}  \\
 a_{s_{2}}   \\
\vdots\\
 a_{s_{N}}  
\end{bmatrix};
  \mathbf{w}(m)=\begin{bmatrix}
{w}_1(m)\\
{w}_2(m)\\
\vdots\\
{w}_N(m)
\end{bmatrix} .
 \end{align*}

The autocorrelation matrix of the received signal in (\ref{RSE}) can be written as 
 \begin{equation}
 \mathbf{R} =\E\{\mathbf{{x}}(m)\mathbf{{x}}(m)^\dag\}=\sigma^2_{\mathrm{s}}\mathbf{a}\mathbf{a}^\dag+{{\sigma_{}}}^2\mathbf{I},
  \label{R}
 \end{equation}
 where $\mathbb{E}\{ s(m)s(m)^*  \} = \sigma^2_\mathrm{s}$ ,  $\mathbb{E}\{\mathbf{w}(m)\mathbf{w}(m)^\dag\} = \sigma^2 \mathbf{I}$, and  $\mathbf{I}$ is an $N\times N$ identity matrix. Since the autocorrelation matrix is symmetric, only the upper triangular part excluding the diagonal terms are collected for training. The angles of these terms are organized in an array $\mathbf{g}$ given by
\begin{multline}
    {\mathbf {g}} = [\angle g_{1,2}, \angle g_{1,3},\ldots, \angle g_{1,N}, \angle g_{2,3},\ldots, \angle g_{2,N},\ldots, \\ \angle g_{(N - 2),(N - 1)}, \angle g_{(N - 2),N}, \angle g_{(N - 1),N}] ,
    \label{z}
\end{multline}
where $g_{i,j} = [\mathbf {R}]_{i,j}, i = 1,\ldots, N-1$, $j= i+1,\ldots, N$. 

To develop the SVR-CFA, no background noise (i.e., ${\sigma}= 0$) is assumed. This assumption yields the phase of $g_{i,j}$ in (\ref{z}) as
  \begin{equation}
\angle g_{i,j}= 2\pi r \sin \theta \left \{\cos(\phi-\beta_i)-\cos(\phi-\beta_j)\right \}/\lambda.
 \label{a_n01}
 \end{equation}
The above assumption will be relaxed in the numerical section. Notice that when the elevation angle, $\theta=0^{\circ}$, all the elements of vector ${\mathbf {g}}$ in (\ref{z}) are same, zero and thus, $\|\mathbf g\| = 0$. If that is the case, one can easily localize the source along $z-$ axis ($\theta= 0^{\circ}$), and the value of $\phi$ has no significance. Throughout the rest of the this work, we will concentrate on the environment, where $0^{\circ}< \theta \leq 90^{\circ}$ and thus $\|\mathbf g\| \neq 0$. Normalizing the vector $\mathbf{g}$, the input data $\mathbf{z}$  for the SVR is obtained as
\begin{equation}
\label{g}
\mathbf{z} = \frac{\mathbf g} {\|\mathbf g\|}\,,
\end{equation}
and the structure of $\mathbf{z}$ will be exploited to prove the following claim, which turns the 2D DOA training and testing  into a 1D SVR problem.

\begin{claim}
For any DOA angle pair $(\theta,\phi)$, $\mathbf{z} \in \Sigma$ (where $\Sigma \subset \mathbb{R}^{N(N-1)/2}$) represents a mapping from the azimuthal angle, $\phi \to \Sigma$, regardless of the elevation angle $\theta$, where $0< \theta \leq 90^{\circ}$. 
\end{claim}
\begin{IEEEproof}
Notice in (\ref{a_n01}) that $u=2\pi r \sin \theta/ \lambda$ is common in every element of $\mathbf{g}$ in (\ref{z}). Since the input vector $\mathbf{z}$  is obtained by normalizing the vector $\mathbf{g}$, this constant, $u$ is cancelled out from every element. Thus, $\mathbf{z}$ will only depend on the variable $\phi$, since $\{\beta_i\}_{i=1}^N$ are constants. Therefore, a mapping is performed from $\phi \to { \Sigma}$ while obtaining $\mathbf{z}$. This establishes the claim.
 \end{IEEEproof}
 
  Our next goal is to find the unknown inverse mapping ${F} {:}\ \Sigma \to \phi$, which will be elaborated upon in the following section. We will then present how the SVR given estimate of $\phi$ will be used to find the elevation angle $\theta$ in a closed form manner.  
 \section{SVR-CFA}
 Let's consider a set of $L$ training pairs which are constructed with $L$ equally spaced fixed angles $\{\phi_i\}_{i=1}^{L}$, and their corresponding output values $\{\mathbf{z}_i\}_{i=1}^{L}$ using (\ref{a_s})-(\ref{g}). From these input/output pairs, our goal is to find a function $\tilde{F}$, which approximates the unknown inverse mapping
function $F$. In \cite{a1, a2}, that function  $\tilde{F}$ is defined by
\begin{equation}
  \hat{\phi}=  \tilde{F} (\mathbf{z}) = \langle {\mathbf {w}}, {\mathbf {\kappa}}(\mathbf {z}) \rangle + b\,,
    \label{p1}
\end{equation}
where $\mathbf {\kappa} (\cdot)$ represents a nonlinear function that transforms the input array
into a high-dimensional space. Now, the SVR parameters, the weight vector $\mathbf {w}$ and the bias term $b$ are obtained by minimizing the regression risk represented by following expression \cite{a1, a3};
\begin{equation}
    R_{\mathrm{reg}}=\frac {1}{2}\| {\mathbf w} \|^{2}+ C\sum \limits _{i=1}^{L} c( \mathbf{z}_i,\phi_i )\,
    \label{reg}.
\end{equation}
Here, $C$ is a constant and $c( \mathbf{z_i},\phi_i )$ is the $\varepsilon$ insensitive loss function, where $\varepsilon$ is the allowed error for the training data~\cite{a0}. This function is defined by  
\begin{equation}
 c( \mathbf{z},\phi ) =  \begin{cases}
0 & | \phi-\tilde{F} (\bf{z})|\leq \varepsilon\\
 | \phi-\tilde{F} (\bf{z})|- \varepsilon & \mbox{otherwise.}
\end{cases}
\end{equation}
Now, the minimization problem of (\ref{reg}) can be solved by standard Lagrange multiplier technique, which then can be written as \cite{a1, a2, a3}
\begin{multline}
    J(\alpha',\alpha)=-\varepsilon\sum \limits _{i=1}^{L} (\alpha'_i+\alpha_i) +\sum \limits _{i=1}^{L} \phi_i(\alpha'_i-\alpha_i) \\
    -\frac{1}{2}\sum \limits _{i=1}^{L} \sum \limits _{j=1}^{L}(\alpha'_i-\alpha_i)(\alpha'_j-\alpha_j)\kappa(\mathbf{z}_i,\mathbf{z}_j),
    \label{l}
\end{multline}
where $\kappa(\mathbf{z}_i,\mathbf{z}_j)$ is obtained using the kernel function which will be defined shortly. The above equation is maximized subject to the constraints $0 \leq \alpha'_i,\alpha_i\leq C$, and $\sum \limits _{i=1}^{L} (\alpha'_i-\alpha_i)=0$, where $\alpha_i$ and $\alpha'_i$ are Lagrange multipliers. This quadratic optimization problem can be solved by using standard quadratic optimization techniques to obtain the weight vector,
\begin{equation}
    \mathbf{w}=\sum \limits _{i=1}^{L} (\alpha'_i-\alpha_i)\kappa(\mathbf{z}_i).
\end{equation}
By substituting $\mathbf{w}$ in (\ref{p1}), we get the estimate of the azimuthal angle,
\begin{equation}
    \hat{\phi}=\tilde{F} (\mathbf{z}) = \sum \limits _{i=1}^{L} (\alpha'_i-\alpha_i)\kappa(\mathbf{z}_i,\mathbf{z}) + b,
    \label{p2}
\end{equation}
where the bias parameter $b$ can be determined as
\begin{equation}
    \hat{b}= \frac{1}{L}\sum \limits _{k=1}^{L}\left\{\sum \limits _{i=1}^{L} (\alpha'_i-\alpha_i)\kappa(\mathbf{z}_i,\mathbf{z}_k)-\phi_k\right\}.
    \label{b}
\end{equation} 
The above estimation of $b$ satisfies the Karush–Kuhn–Tucker (KKT) conditions mentioned in \cite{a2}, since $\eta_1\leq \hat{b} \leq \eta_2$, where 
$$ \eta_1 = \rm{max} \left \{  - \varepsilon + \phi_k - \langle {\mathbf {w}}, {\mathbf {\kappa}}(\mathbf{z}_k) \rangle\hspace{1.5 mm} |\hspace{1.5 mm} \alpha'_k< C \hspace{1.5 mm}\rm{or}\hspace{1.5 mm}  \alpha_k>0 \rangle  \right \}, $$
\vspace{-15pt}
$$\eta_2= \rm{min} \left \{  - \varepsilon + \phi_k - \langle {\mathbf {w}}, {\mathbf {\kappa}} (\mathbf {z}_k) \rangle \hspace{1.5 mm} | \hspace{1.5 mm} \alpha'_k> 0 \hspace{1.5 mm} \rm{or} \hspace{1.5 mm} \alpha_k< C \rangle  \right \},$$ and $k=1,\ldots, L$. Finally, we use the estimate of the azimuthal angle, $\phi$ from (\ref{p2}) with $\{\angle g_{i,j}\}$ of  (\ref{a_n01}) and average the normalized version of those terms to obtain the elevation angle as
    \begin{equation}
    \hat{\theta}= \arcsin\left\{\frac{\lambda}{2\pi r}\times \mathrm{avg}\paren{ \frac{ \angle g_{i,j}}{\gamma_{i,j}} }
 \right\},
   \label{th}
    \end{equation}
 where $\gamma_{i,j}=\cos(\phi-\beta_i)-\cos(\phi-\beta_j)$. Note that $\gamma_{i,j}$ are selected in such a way that $\gamma_{i,j} \neq 0$. Next, we move on to the numerical study of this work.

\section{Numerical Study}
\label{Na}
Here, we first describe the implementation of the proposed SVR-CFA, and then demonstrate its performance using the MUSIC as the baseline algorithm.
\vspace{-10pt}
\subsection{Implementation of the SVR-CFA}
We consider a front-looking radar with a field of view (FOV): elevation angle $\theta \in [1^{\circ},90^{\circ}]$, and azimuthal angle $\phi \in [0^{\circ},180^{\circ}]$. We choose the inter element spacing $d = \lambda/2$. For the training purpose, we consider the elevation angle, $\theta = 30^{\circ}$, and the azimuthal angle, $\phi$ is varied from 0 to $180^{\circ}$ with $1^{\circ}$ angular separation. Using (\ref{a_s})-(\ref{g}), we obtain $L = 181$ training vectors $\{\mathbf{z}_i\}$ corresponding to these azimuthal angles considering an extremely large signal to ratio (SNR) environment. We choose constant $C=\rm{max}(|\mu+3\sigma_{\phi}|,|\mu-3\sigma_{\phi}|)$, where $\mu$, and $\sigma_{\phi}$ are the mean and the standard deviation of the azimuthal angle of the training set \cite{a7}, and we find, $C= 247.18$. 

The nonlinear transformation (i.e., $\mathbf{\kappa}$) used in this work is  a radial basis kernel function \cite{a2, a6}, defined as
\begin{equation}
   \bf{ \kappa} (\bf{z}_{i},\bf{z}) = \exp ({-} {\delta}{\Vert \bf{z} - \bf{z}_{i}\Vert^{2}}),
\end{equation}
where the kernel width parameter $\delta$ reflects the input range of the training/test data. Therefore, this RBF kernel represents this similarity as a decaying function of the distance between the training/testing vectors. In this work, we use $\delta=0.5$, and as per the suggestion in \cite{a8}, we set the allowed error, $\varepsilon=1.0043\,{\sigma}$. Now, the dual optimization problem in (\ref{l}) is solved for the unknown coefficients $\alpha'_i$, and $\alpha_i$ using ``CVX (MATLAB)", and the weight vector, $\mathbf{w}$ is obtained. The bias term $b$ in (\ref{p1}) is then calculated using (\ref{b}).

For the testing, we consider a different elevation angle, $\theta = 60.5^{\circ}$, and the azimuthal angle, $\phi$ is varied from $0.5^{\circ}$ to $180.5^{\circ}$ with $1^{\circ}$ angular separation so that it contains angles that are not included in the training phase. Detection of these angles will ensure the generalization capability of the SVR. We estimate the autocorrelation matrix $\mathbf{R}$ by averaging $M=30$ snapshots using  
\begin{equation}
    \hat{\mathbf{R}}=\frac{1}{M}\sum_{m=1}^M \mathbf{{x}}(m)\mathbf{{x}}^\dag(m).
\end{equation}
Now, we obtain the SVR input vector, $\bf{z}$ using (\ref{g}) from the elements of matrix $\hat{\mathbf{R}}$. The azimuthal angle $\phi$, and the elevation angle $\theta$ can be estimated using (\ref{p1}), and (\ref{th}), respectively. Note that while obtaining the phase angles, ambiguity may occur which can be resolved utilizing the techniques suggested in \cite{zoltowski1994real, mod1}. 

\subsection{SVR-CFA vs MUSIC}
 In practical applications, not only the DOA estimation accuracy is important but also the computational complexity plays a significant role. In this work, the complexities of the algorithms are measured by the number of real-time multiplications associated with the major operations. Note that one complex multiplication is equivalent to four real multiplications.
\begin{table}[t]
\begin{center}
\caption{Complexity analysis}
\vspace{-2pt}
\begin{tabular}{|c|c|}
\hline
\textbf{Algorithm}&\textbf{Complexity (number of real multiplications)}\\
\hline
\multirow{2}{*}{MUSIC} & $48N^3+4N^2M+2N^2$  \\
  & $+N_{\theta}N_{\phi}\{4N^2-4N+3\}$\\
\hline
\multirow{2}{*}{SVR-CFA} & $2NM(N-1)+3N(N-1)+L^2+2$  \\
  & $+\log_2 p\times\{ 3N(N-1)/2+1\} $\\
\hline
\end{tabular}
\label{t0}
\end{center}
\end{table}

\begin{figure}[t!]
\centering
    \includegraphics[scale=0.40]{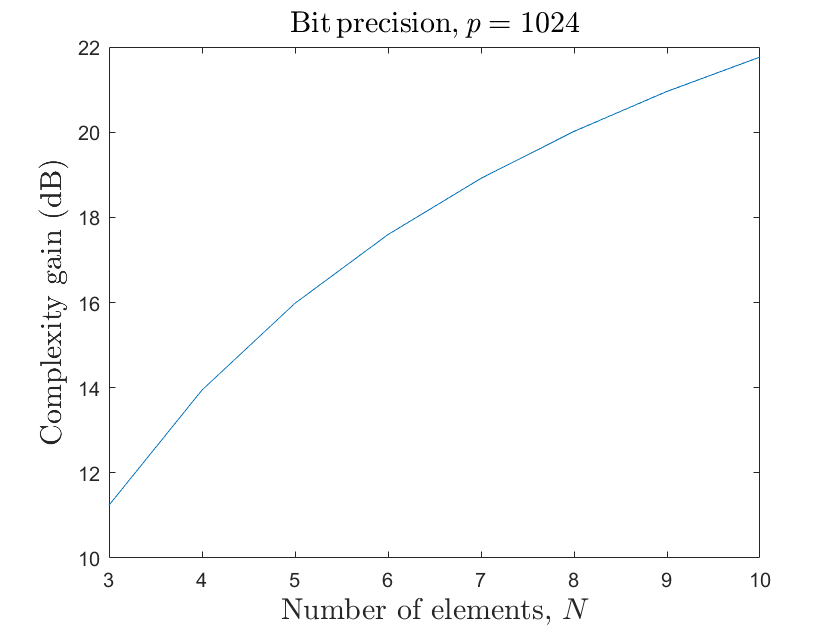}
    \caption{Complexity gain of SVR-CFA over MUSIC.}
    \label{figg}
\end{figure}
To implement MUSIC, the sample autocorrelation matrix is needed, and requires $4N^2M+2N^2$ real multiplications. On this matrix, MUSIC performs eigen-value decomposition (EVD), which often is obtained from a singular value decomposition (SVD). As per \cite{evd}, this complexity associated with an SVD is $48N^3$. For the 2D MUSIC, the cost of the DOA angle search using the null space requires $N_{\theta}N_{\phi}\{4(N-1)^2+2(N-1)+2N+1\}$ real multiplications, where $N_{\theta}$ and $N_{\phi}$ represent the searching point number on the azimuthal and elevation planes, respectively. Recall that the SVR-CFA only requires the non diagonal upper half of the autocorrelation matrix. The associated cost is $2NM(N-1) +N(N-1)$ real multiplications. After that the SVR-CFA calculates the normalized phase with a complexity of $N(N-1)/2+\log_2 p \times N(N-1)/2+N(N-1)+1$, where $p$ refers to the number of digits of precision \cite{cost,bc}. Here, the first two terms are related to the phase extraction ($\ref{z}$), and the remaining terms are associated with the normalization of the SVR input vector ($\ref{g}$). In addition, the testing part of the SVR-CFA requires $L^2$ real multiplication to estimate the azimuthal angle $\phi$, where $L$ is the number of training data. Lastly, to estimate the elevation angle $\theta$ using (\ref{th}), the proposed algorithm requires an additional cost of $\log_2 p\times N(N-1)+ N(N-1)/2+1+\log_2 p$ real multiplications. Here, the first term refers to the cost to get the normalizing terms $\{\gamma_{i,j}\}$, the middle two terms are for the maximum number of divisions and multiplications, and the last term is to calculate the arcsine defined in (\ref{th}). All the corresponding costs are added and displayed in Table \ref{t0} for complexity comparison. According to this table, the cubic order of the array size, and the product of the searching points dominate the complexity of the MUSIC algorithm. On the other hand, the complexity of the SVR-CFA is dictated by the square of the array size, and the length of the training sequence.

Now, we use Table \ref{t0} to compute the complexity gain of the SVR-CFA over the MUSIC algorithm; see  Fig. \ref{figg}. Note that in a single user environment, to estimate the DOA (azimuthal and elevation) angles unambiguously, the minimum value of $N$ is 3. In Fig. \ref{figg}, the search in MUSIC is conducted with $1^{\circ}$ precision of the DOA angles. Here, it can be noticed that the SVR-CFA always shows significant complexity gain over the MUSIC and as expected, this gain increases with the number of antenna elements $N$. At $N=3$, the complexity gain is 11.24 dB and increases to 21.76 dB at $N=10$.   

\begin{figure}[t!]

\begin{minipage}[b]{0.45\linewidth}
\centering
\includegraphics[width=135 pt]{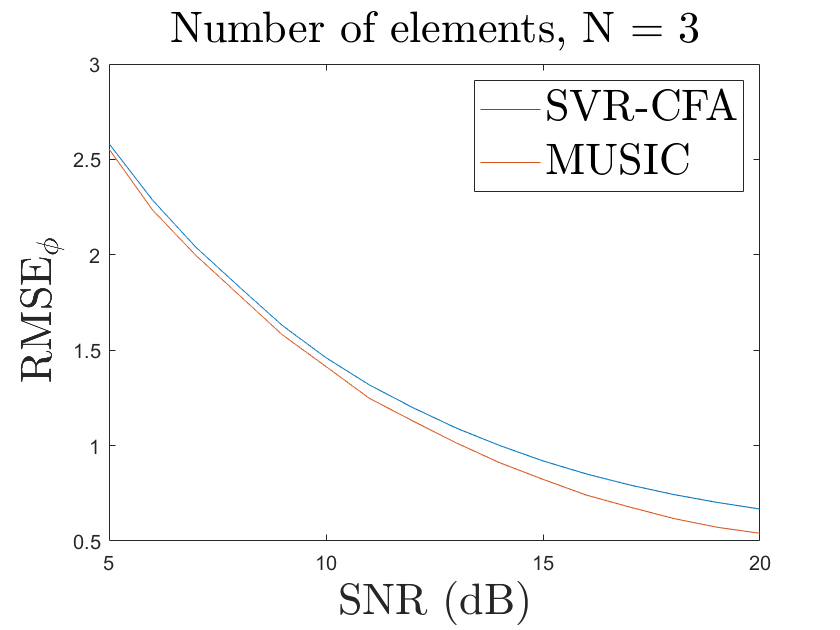}
\caption{{RMSE}$_\phi$ vs SNR.}
\label{N=32}
\end{minipage}
\hspace{0.5cm}
\begin{minipage}[b]{0.47\linewidth}
\centering
\includegraphics[width=135pt]{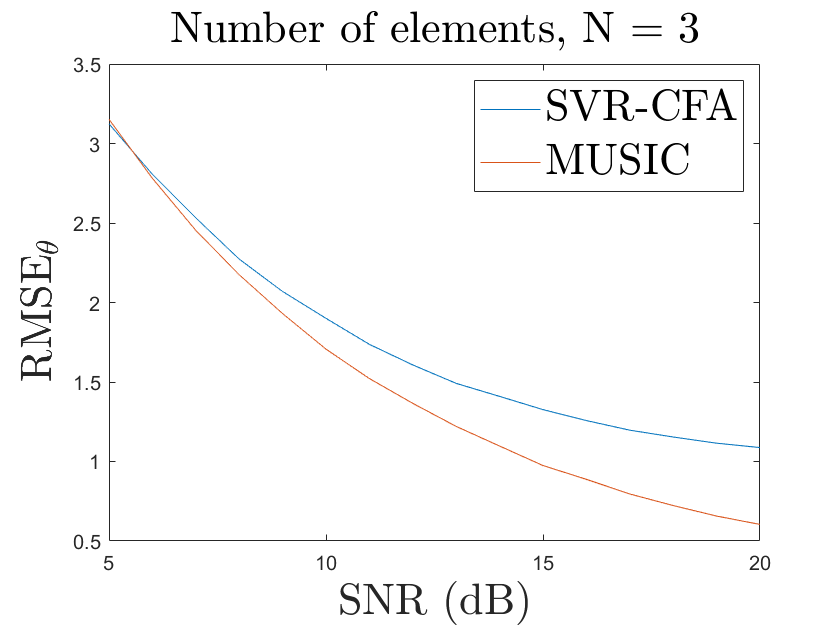}
\caption{{RMSE}$_\theta$ vs SNR.}
\label{N=31}
\end{minipage}
\end{figure}

RMSE is a commonly used metric to measure the estimation accuracy of DOA algorithms. First, we present the RMSEs of both the algorithms as a function of SNR using the simplest array configuration ($N=3$); see Fig. \ref{N=32}, and \ref{N=31}. Here, it can be noticed that the estimation accuracy of the SVR-CFA is similar to the MUSIC algorithm on both the azimuthal and elevation planes specially at low SNR (less than 10 dB). For example, at SNR = 10 dB, RMSE losses on the elevation, and azimuthal planes are $0.05^{\circ}$, and $0.19^{\circ}$, respectively. At high SNR (greater than 15 dB), MUSIC offers sightly smaller RMSE than the SVR-CFA at the cost of 11.24 dB higher complexity.

\begin{figure}[t!]
\begin{minipage}[b]{0.45\linewidth}
\centering
\includegraphics[width=135 pt]{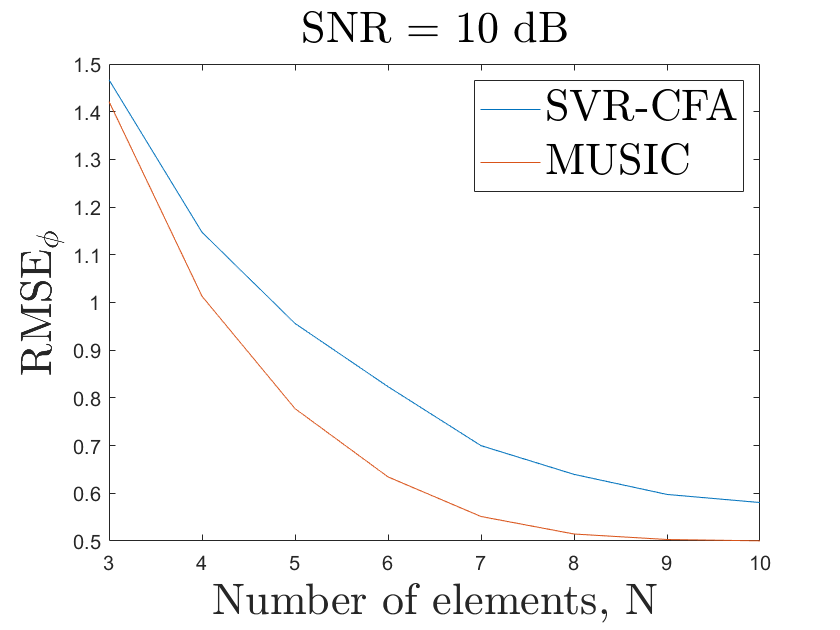}
\caption{{RMSE}$_\phi$ vs $N$.}
\label{N4}
\end{minipage}
\hspace{0.5cm}
\begin{minipage}[b]{0.45\linewidth}
\centering
\includegraphics[width=135pt]{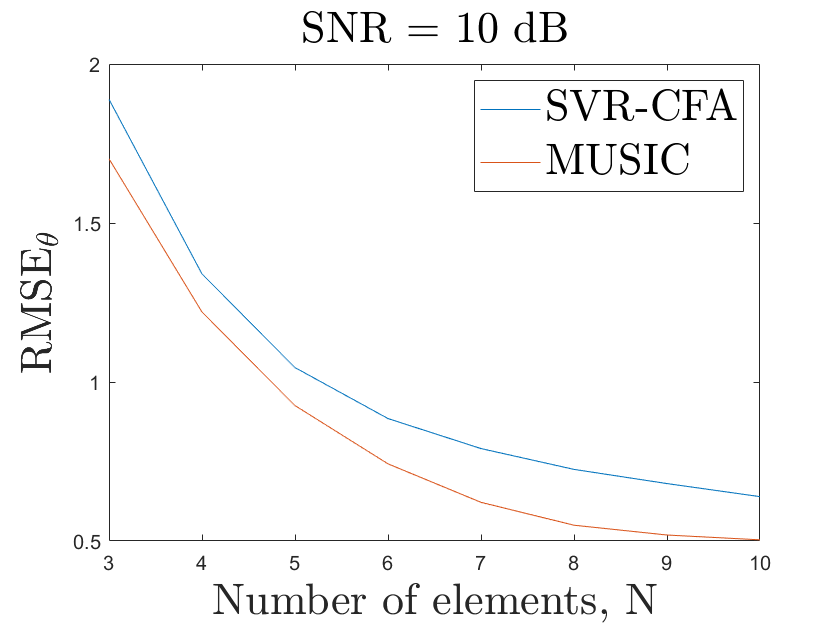}
\caption{{RMSE}$_\theta$ vs $N$.}
\label{N3}
\end{minipage}
\end{figure}
Finally, we compare RMSEs of both the algorithms by varying the number of antenna elements $N$, for SNR  = 10 dB; see  Fig. \ref{N4}, and \ref{N3}. Here, it can be noticed that as expected, RMSE decreases with the increase of $N$ for both the algorithms. Unsurprisingly, MUSIC is slightly more efficient than the SVR-CFA in exploiting the increase of the number of antenna elements in the array. As demonstrated, this gain comes at the cost of rapid surge in the complexity burden.  

\vspace{-10pt}
\section{Conclusions}
In this correspondence item, we addressed the use of a sequential DOA estimation technique, namely SVR-CFA, which combines the reduced dimension SVR with a closed form approach. Using the upper half of the autocorrelation matrix, the input sequence for the SVR is derived in such a way that it is independent of the source's elevation angle. Thus, the SVR requires only 1D implementation. The proposed approach offers significant complexity gain over the popular MUSIC algorithm with negligible performance loss. For instance, in a UCA with $N=3$, the complexity gain is 11.24 dB while exhibiting $0.05^{\circ}$, and $0.19^{\circ}$ precision loss on the elevation, and the azimuthal planes, respectively, at SNR = 10 dB.

 \bibliographystyle{IEEEtran}
\bibliography{Reference.bib}

\end{document}